\documentclass[fleqn]{rasti}

\usepackage{lmodern}
\usepackage{printlen}
\usepackage{listings}
\usepackage{newtxtext,newtxmath}
\usepackage{siunitx}
\usepackage[backend=biber, style=ext-authoryear]{biblatex}
\let\cite\parencite

\usepackage[T1]{fontenc}
\usepackage{tikz}
\usetikzlibrary{shapes,arrows,positioning, fit, backgrounds, calc}

 \lstdefinelanguage{stimela}{
    basicstyle=\ttfamily\footnotesize,
    morekeywords={info,inputs, outputs, assign, steps, cab, skip, params, opts, _include},
    sensitive=false,
    morecomment=[l]{\#},
    morecomment=[s]{/*}{*/},
    morestring=[b]",
    string=[s]{'}{'},
    stringstyle=\color{blue},
}

 \DeclareSIUnit \Jansky {\ensuremath{\mathrm{Jy}}}

\DeclareRobustCommand{\VAN}[3]{#2}
\let\VANthebibliography\thebibliography
\def\thebibliography{\DeclareRobustCommand{\VAN}[3]{##3}\VANthebibliography}

\usepackage{graphicx}	
\usepackage{amsmath}	
\usepackage{circuitikz}
\usepackage{caption}
\usepackage{subcaption}

\newcommand{\orcid}[1]{\href{https://orcid.org/#1}{\includegraphics[width=8pt]{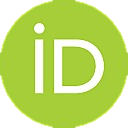}}}
\graphicspath{{img/}}

\title[TART interferometer]{The Transient Array Radio Telescope: An open-source all-sky imaging radio interferometer}

\author[Molteno et al.]{
Timothy C.A. Molteno \orcid{0000-0002-2022-0380}$^{1,2,3}$\thanks{E-mail: tim@physics.otago.ac.nz (TCAM)}
Max Scheel$^{2}$,
Pat Suggate$^{2}$,
Ben V. Hugo \orcid{0000-0002-2933-9134}$^{4}$,
Phil Brown$^{1}$
\\
$^{1}$ Department of Physics, University of Otago, Dunedin, New Zealand\\
$^{2}$ Electronics Research Foundation, Dunedin, New Zealand \\
$^{3}$ Department of Physics \& Electronics, Rhodes University, South Africa \\
$^{4}$ South African Radio Astronomy Observatory (SARAO), WC, Cape Town, South Africa.}

\date{Accepted XXX. Received YYY; in original form ZZZ}

\pubyear{2026}

\begin{document}
\label{firstpage}
\pagerange{\pageref{firstpage}--\pageref{lastpage}}
\maketitle

\begin{abstract}
The Transient Array Radio Telescope (TART) is an open-source 24-element 1573-1577 \unit{\mega\hertz} L-band imaging interferometer designed for research and education. This paper introduces the physical architecture of the TART, describes it's electronics, radios, correlator, and the design of the antenna array. Then we discuss the software architecture, including the calibration of the telescope using satellites and imaging pipelines. Imaging results are presented, which lead to an estimate of $\sim52 \unit{\kilo\Jansky}$ for the sensitivity of the telescope. 

\end{abstract}




\section{Introduction}

Many large astronomical radio telescopes operate on the principle of aperture-synthesis~\cite{thompson2017interferometry}, which allows an image of the radio sky to be produced from a number of fixed receivers. First developed in the late 1950s~\cite{ryle1960synthesis, ryle1962radio, swenson1969synthetic}, aperture-synthesis telescopes can achieve extraordinary sensitivity~\cite{mauch20201} and angular resolution~\cite{event2019first}. Typically these aperture synthesis telescopes are large and very costly instruments that are designed and custom-engineered by national~\cite{napier1983very, swarup1991giant, jonas2016meerkat, johnston2008science} or, in the case of the SKA, multi-national~\cite{carilli2004science} teams.


The Transient Array Radio Telescope (TART) project described in this paper uses modern, highly integrated electronic components to realize a wide-field, aperture synthesis, open-source radio telescope. Operating with 24 separate radio receivers at 1575~\unit{\mega\hertz} with a 2.5~\unit{\mega\hertz} bandwidth, the system is engineered to cover the entire visible sky continuously. Modern time-domain astrophysics and space situational awareness increasingly rely on high-cadence, hemispherical monitoring to capture bright, fast-fading, or unpredictable transient events without the latency and narrow field-of-view constraints of mechanically steered dishes. 


An advantage of operating in the GNSS L1-band is that significant Radio Frequency Interference (RFI) is unusual, as it would limit the functionality of global navigation systems and mobile phone networks. Transients, such as lightning, while not uncommon, are of interest to observers, and the detection of such short transients is one of the goals of the TART. The GNSS satellite signals themselves, at $\sim 60  \unit{\mega\Jansky}$, are below the noise-floor of the TART receiver and therefore do not saturate the receivers. These signals are, in-fact, required for continuous calibration of the TART. Practical experience with TART data indicates that RFI is rare --  present in fewer than one percent of snapshot observations (one second integrations). We can therefore  operate our imaging pipeline without the need to flag RFI.

The architecture of TART is optimized to detect transient radio phenomena with an all-sky field-of-view and continuous full-sky synthesis imaging -- ideal for capturing spontaneous, high-flux atmospheric interactions. The primary science goal for TART are radio signatures generated by high-energy cosmic ray (HECR) interactions with the Earth's upper atmosphere.  Beyond cosmic ray interactions, TART's unsteered, wide-field design operating at 1575~\unit{\mega\hertz} enables non-astronomical applications, including space situational awareness, satellite tracking, and monitoring near-Earth objects (NEOs) emitting or reflecting signals in the L1 band. TART prioritizes field of view over sensitivity; this all-sky coverage limits the maximum gain of its individual antenna elements to 3$~\mathrm{dBi}$~\cite{kraus2002antennas} which allows compact individual antennas without sacrificing telescope performance.

\begin{figure*}
\includegraphics[width=\linewidth]{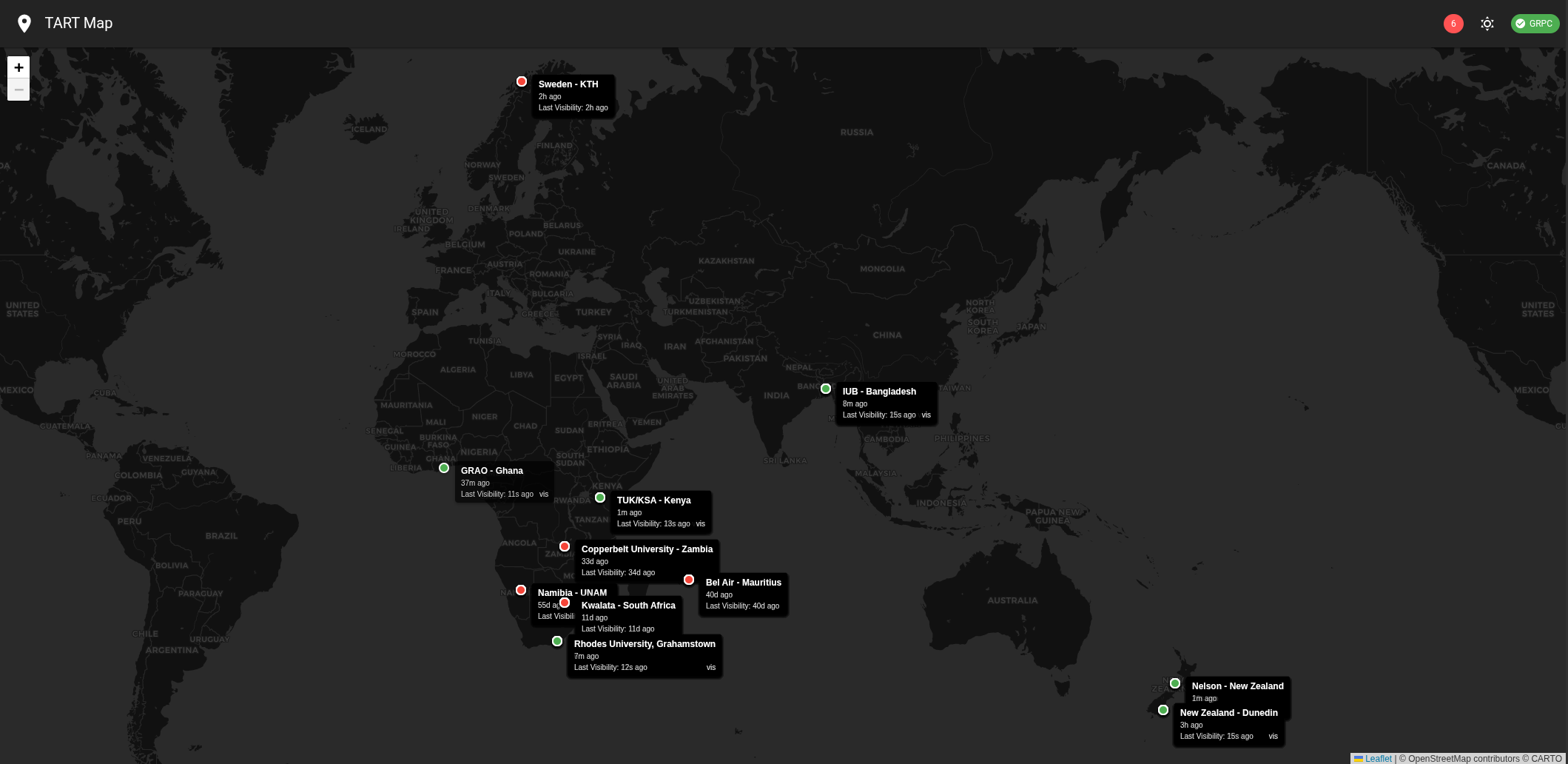}
\caption{The online TART world map (\url{https://map.elec.ac.nz}) as of April 2026. This snapshot shows the current TART installations around the globe.}
\label{fig:tart_web_map}
\end{figure*}

\begin{figure}
\includegraphics[width=\linewidth]{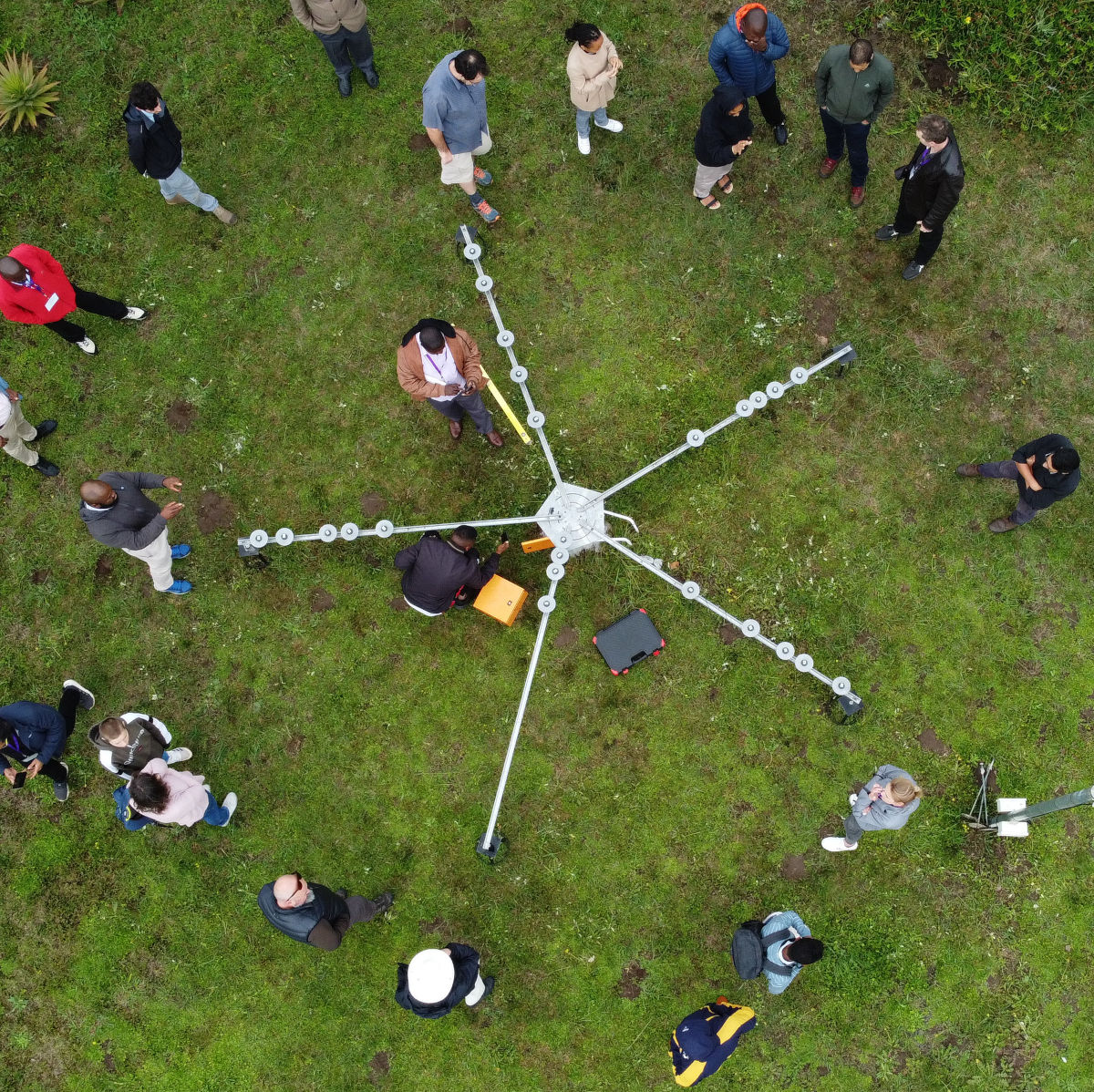}
\caption{Aerial view of the TART telescope installed at Rhodes University, South Africa. The antennas on this telescope are laid out in a five-arm design, designed to allow easy movement of antennas along each arm. Using this design, the effects of modifying antenna positions on the interferometer response can easily be studied.}
\label{fig:tart_y_array}
\end{figure}

The result is an open-source, 24-element all-sky radio telescope that serves as a detector for bright transients and orbital targets while also providing an accessible testbed for algorithm development and hands-on radio astronomy instrumentation. To date, eleven TART systems have been deployed globally (Figure~\ref{fig:tart_web_map}), originating in New Zealand, expanding in eight African countries, and extending into Asia and Europe. Figure~\ref{fig:tart_y_array} shows an aerial view of a TART installation configured in a five-armed array layout, with the core electronics housed at the central hub.



In Section~\ref{sec:hardware} we introduce TART hardware, RF electronics, antennas, array layout, and the real-time correlator. In Section~\ref{sec:software}, we describe the components of the TART software, including the operating software and imaging pipelines. Section~\ref{sec:tart_imaging_results} presents TART imaging results from which a telescope sensitivity estimate is obtained. Finally, in Section~\ref{sec:discussion} we conclude with a discussion of some future work planned for the project.

\section{Hardware Architecture}
\label{sec:hardware}

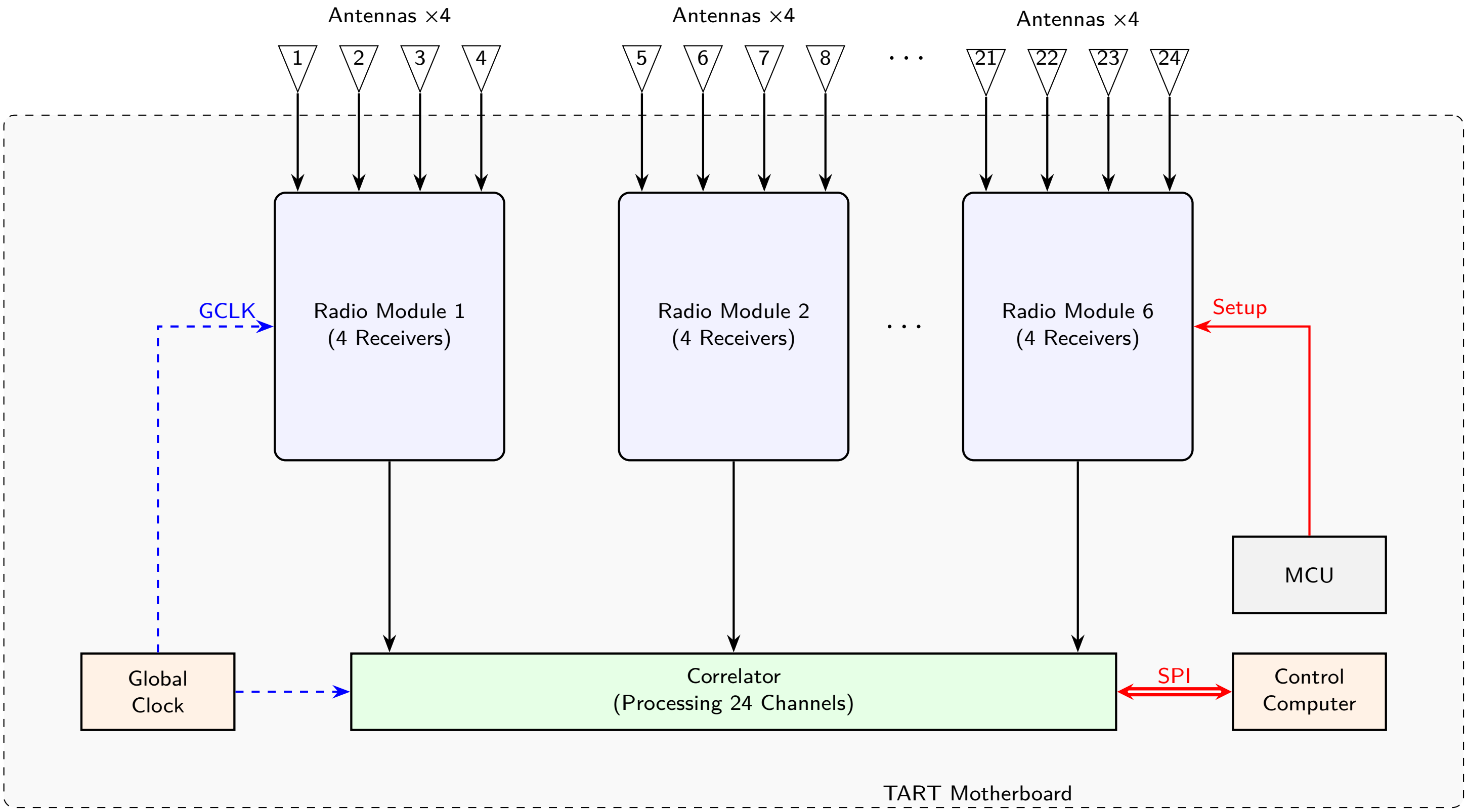
\begin{figure*}
\begin{tikzpicture}[
    node distance=1.5cm,
    font=\sffamily\small,
    antenna/.style={isosceles triangle, draw, shape border rotate=-90, minimum width=0.5cm, minimum height=0.5cm, inner sep=0},
    module/.style={rectangle, draw, fill=blue!5, minimum width=3cm, minimum height=3.5cm, rounded corners, align=center, thick},
    component/.style={rectangle, draw, fill=orange!10, minimum width=2cm, minimum height=1cm, align=center, thick},
    mb/.style={rectangle, draw, dashed, inner sep=1cm, fill=gray!5, rounded corners},
    spi_bus/.style={<->, >=stealth, very thick, double, double distance=1pt},
    arrow/.style={-Stealth, thick}
]

    \foreach \m [count=\i] in {1, 2, 6} {
        \node[module] (mod\i) at ({(\i-1)*4.5}, 0) {Radio Module \m \\ (4 Receivers)};
        
        \ifnum \i=2
            \node at ($(mod\i.east) + (0.75, 0)$) {\LARGE $\dots$};
        \fi

        \foreach \a in {1, 2, 3, 4} {
            \node[antenna] (ant\i\a) at ($(mod\i.north) + ({-1.2 + (\a-1)*0.8}, 1.75)$) {\pgfmathparse{int(\a+(\m-1)*4)}\pgfmathresult};
            \draw[arrow] (ant\i\a.south) -- ($(mod\i.north) + ({-1.2 + (\a-1)*0.8}, 0)$);
        }
        
        \node[above=0.2cm of ant\i2, xshift=0.4cm] {\small Antennas $\times$4};

        \ifnum \i=2
            \node at ($(ant\i2.east) + (2.5, 0)$) {\LARGE $\dots$};
        \fi

    }

    \node[component, minimum width=10cm, fill=green!10] (correlator) [below=2.5cm of mod2] {Correlator \\ (Processing 24 Channels)};

    \draw[arrow] (mod1.south) -- (mod1.south |- correlator.north);
    \draw[arrow] (mod2.south) -- (mod2.south |- correlator.north);
    \draw[arrow] (mod3.south) -- (mod3.south |- correlator.north);

    \node[component] (clock) [left=1.5cm of correlator] {Global \\ Clock};
    \node[component] (computer) [right=1.5cm of correlator] {Control \\ Computer};
    \node[component, fill=gray!10] (mcu) [above=0.5cm of computer] {MCU};

    \draw[arrow, dashed, blue] (clock.east) -- (correlator.west);
    \draw[arrow, dashed, blue] (clock.north) |- (mod1.west) node[pos=0.8, above] {GCLK};
    
    \draw[spi_bus, red] (computer.west) -- node[above] {SPI} (correlator.east);
    \draw[arrow, red] (mcu.north) |- (mod3.east) node[pos=0.8, above] {Setup};

    \begin{scope}[on background layer]
        \node[mb, fit=(mod1) (mod3) (correlator) (clock) (computer), label={[anchor=south east]below right:TART Motherboard}] (motherboard) {};
    \end{scope}
\end{tikzpicture}
    \caption{Block diagram of components of the TART telescope. Each of the 24 antennas connects to a receiver located on a radio module that contains four radio receivers. The receivers in turn send data to the correlator. Each receiver is synchronously clocked using the global clock. There are six radio boards for a total of 24 receivers in the TART interferometer. Correlation products or raw data are read over a Serial Peripheral Interface (SPI) bus by the control computer. An independent microcontroller unit (MCU) is used to initially configure all radios.}
    \label{fig:tart_block_diagram}
\end{figure*}

\begin{figure}
\includegraphics[width=\linewidth]{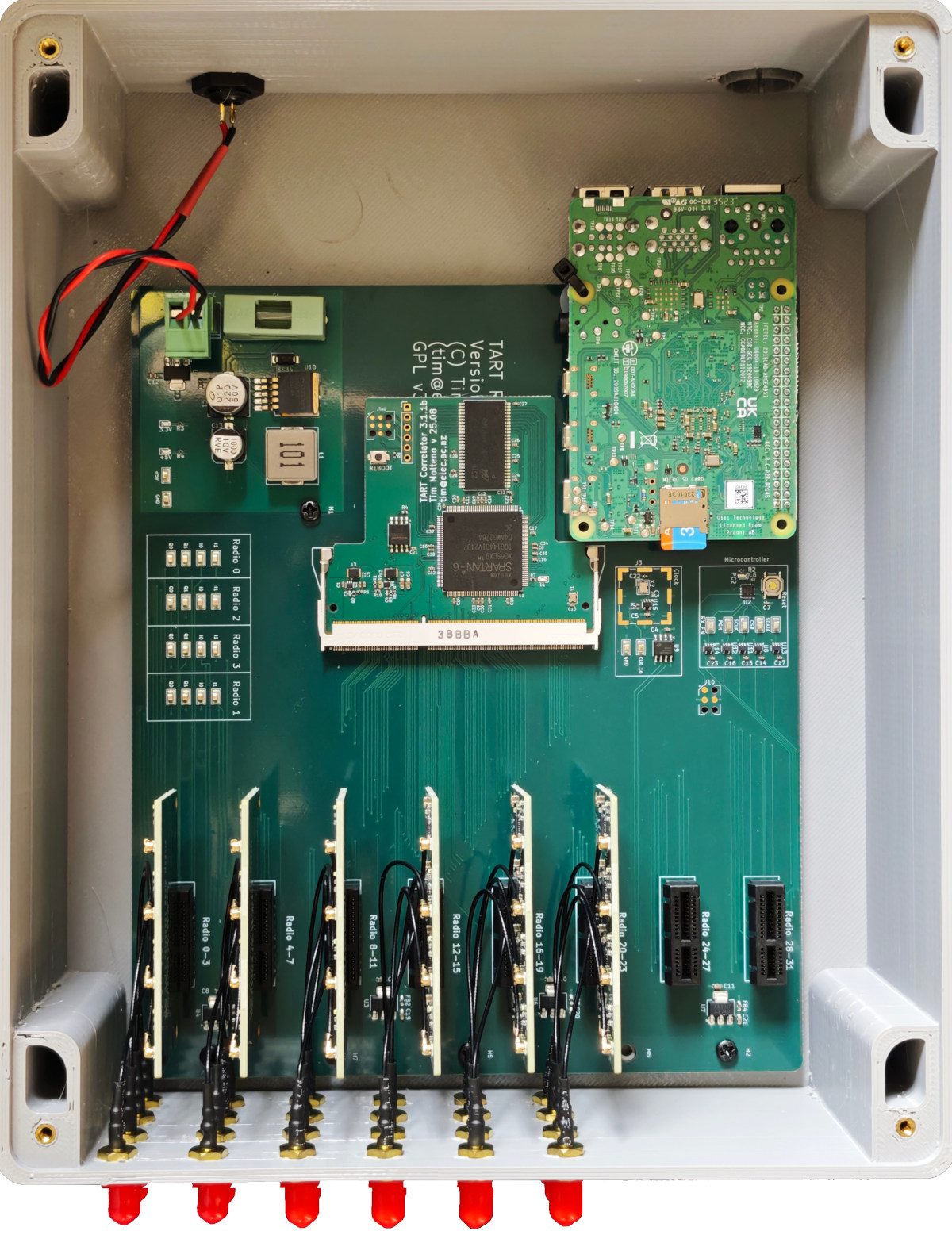}
\caption{TART-3 electronics development kit with 24 radio receivers installed. The control computer is at the top-right, the correlator is mounted in a socket in the center and six radio boards, each with four radio receivers are located on the bottom. There are two empty slots that can be populated to increase the number of antennas from 24 to 32. The 24 SMA connectors are covered in red protective covers in the image. }
\label{fig:development_kit}
\end{figure}

The TART telescope consists of four primary elements; antennas, radio receivers, the local oscillator, and the signal correlator. In addition, there is a telescope control computer that manages the state of the telescope and provides an API for accessing telescope data and controlling the telescope. The general block diagram is shown in Figure~\ref{fig:tart_block_diagram}, and shows the principal components of the TART. A central reference clock at 16.368\unit{\mega\hertz} is distributed to all receivers and the correlator and provides a phase reference, ensuring that the data from all receivers are sampled synchronously. Figure~\ref{fig:development_kit} shows the physical layout of the electronics.

\subsection{Antennas and Antenna Array}

Antennas for the TART are typically commercial off-the-shelf Global Navigation Satellite System (GNSS) antennas. These are widely available and include a low-noise amplifier to overcome cable losses to the input signal for each radio. These antennas are typically electrically small patch-antennas designed to receive Right-Hand Circularly Polarized signals with near-hemispherical beam patterns.

The TART antenna array design is a trade-off between large baselines -- giving high angular-resolution, and compactness -- allowing ease of installation and reducing bandwidth-smearing.



\begin{figure}
\includegraphics[width=\linewidth]{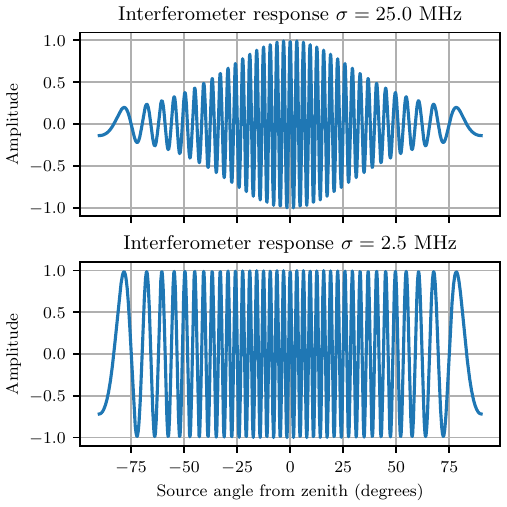}
\caption{Point source interferometer response for the TART telescope with a baseline $D=3.5\unit{\meter}$, for two different bandwidths $\sigma = 25$MHz in the upper plot, and $\sigma=2.5$ MHz in the lower plot. The lower plot shows that the white-light fringes cover the entire field of view of the TART for baselines less than 3.5 meters.}
\label{fig:point-source-interferometer-response}
\end{figure}

By limiting the maximum baseline length, the TART telescope can operate in a single-frequency mode that reduces the load on the correlator. Following~\cite[Equation 2.10]{thompson2017interferometry}, the finite-bandwidth interferometer-response as a function of geometric delay, $\tau_g$, is given by
\begin{equation} 
r(\tau_g)= \exp\left[  -2 \left( \frac{\pi D \sigma}{c} \sin(\theta) \right)^2 \right] \cos \left( \frac{2 \pi D \nu_0}{c} \sin(\theta) \right), 
\end{equation}
where $\sigma$ is the bandwidth, $D$ the baseline, and $\theta$ the angle of the point source from the center of the phase (zenith for the TART). The TART thus exhibits bandwidth smearing, reducing fringe contrast away from zenith. The interferometer response is shown in Figure~\ref{fig:point-source-interferometer-response} for two different bandwidths. We can see that for full-sky response the fringes will be visible at all angles for $\sigma = 2.5  \unit{\mega\hertz}$, but for higher bandwidth ($\sigma = 25 \unit{\mega\hertz}$, the fringe amplitude will be significantly reduced away from the zenith. 

The fringe-washing function, $\alpha(\theta) = \exp\left[  -2 \left( \frac{\pi D \sigma}{c} \sin(\theta) \right)^2 \right]$, provides the envelope of the fringe amplitude as a function of $\theta$. At the edge of the TART field-of-view, $\theta = \frac{\pi}{2}$, this becomes;
\[ \alpha_{min} \equiv \alpha \left( \frac{\pi}{2} \right) = \exp\left[  -2 \left( \frac{\pi D \sigma}{c} \right)^2 \right]. \]
This can be rearranged to get an expression for the maximum baseline, $D_{max}$ in terms of bandwidth, $\sigma$, and a minimum fringe-washing, $\alpha_{min}$,
\[ D_{max} =  \sqrt{-\frac{\ln(\alpha_{min})}{2}} \frac{c}{\pi \sigma}. \]
Assuming a bandwidth $\sigma=2.5 \unit{\mega\hertz}$ and that the fringe amplitude is degraded by 10\% at the edge of the field of view ($\alpha = 0.9$), we get a maximum TART baseline of ~17.5 meters. Thus, for full-bandwidth all-sky imaging, the TART antenna  array must contain baselines less than ~17.5 meters\footnote{Note that this derivation assumes a Gaussian bandpass-response, whereas the TART has a Butterworth 5th-order bandpass-response (see Section~\ref{sec:radio_front_end}).}. For convenience, we have chosen to limit TART baselines to below 4 meters to lower cost and make transport and installation easier. 

\subsubsection{Antenna array layout}

\begin{figure*}
\includegraphics[width=\linewidth]{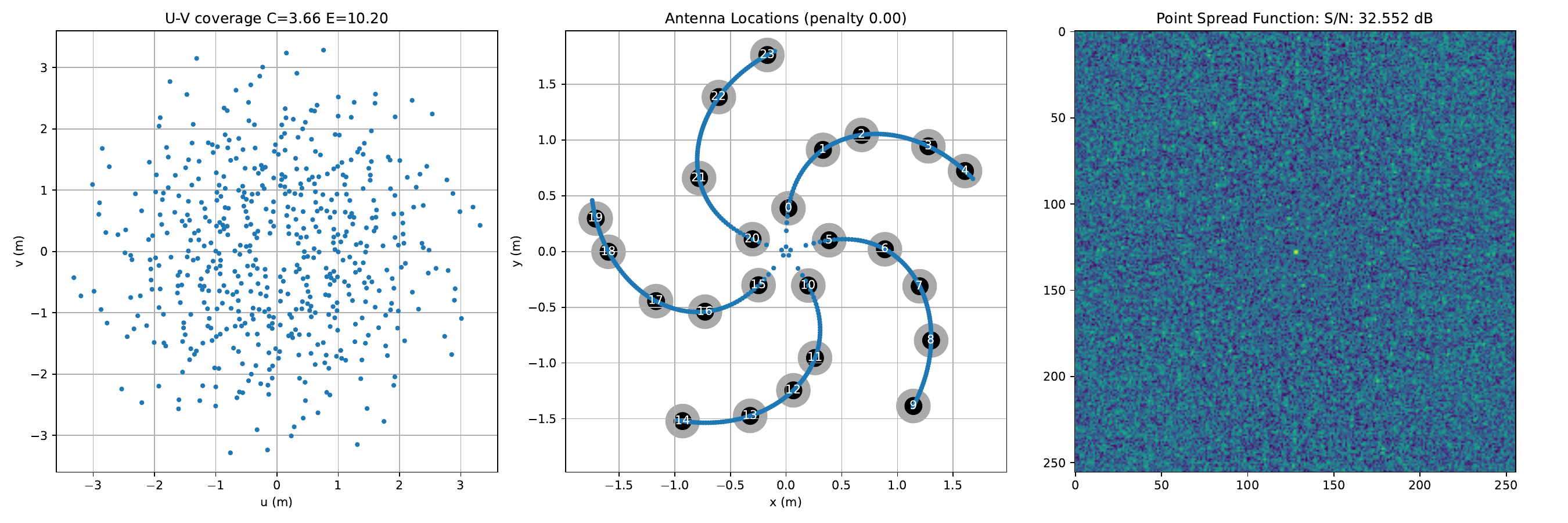}
\caption{A five arm spiral antenna layout for the TART antenna array. The figure on left shows the $u-v$ coverage of the array, in the middle is the physical layout of the array, and on the right is the point spread function.  These arms are Fermat Spirals.}
\label{fig:tart_spiral_array}
\end{figure*}

Array layouts are designed using an optimization strategy that minimizes side lobes in the point-spread function~\cite{thompson2017interferometry}, while keeping antenna separation above a lower threshold to minimize antenna coupling. To date, TART deployments have used two types of antenna arms for the telescope array. Straight arm arrays as shown in Figure~\ref{fig:tart_y_array} are ideal for easy modification and measurement of antenna locations. More compact spiral arrays, such as those shown in Figure~\ref{fig:tart_spiral_array}, provide improved performance in a more compact space.  

\subsection{Radio Front End}
\label{sec:radio_front_end}

\begin{figure}
    \includegraphics[width=\linewidth]{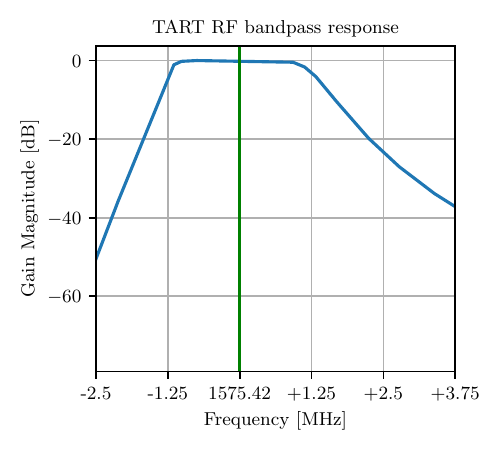}
    \caption{Frequency response of the RF front-end, configured to 2.5 \unit{\mega\hertz} bandwidth with a 5th order Butterworth bandpass IF filter.}
    \label{fig:rf_bandpass_response}
\end{figure}

Each TART receiver uses a highly integrated GNSS radio front-end designed to digitize Radio Frequency (RF) signals in the L1-band~\cite{max2769b:datasheet}. These integrated receivers are widely used in commercial GNSS positioning devices and have been optimized to have good low-noise performance. The receiver is configured to operate with a 2.5 \unit{\mega\hertz} bandwidth centered about 1575.42 \unit{\mega\hertz} (see Figure~\ref{fig:rf_bandpass_response}). This configuration is performed for all radios on startup by a micro-controller located on the TART motherboard.

\subsection{Correlator}

The TART telescope correlator performs real-time correlation of all visibilities. Single-bit I and Q data are streamed in parallel from each radio front end into a Field Programmable Gate Array (FPGA). As the TART operates in single-frequency mode, the correlator just performs cross-correlation between each pair of antennas over an integration-period $\tau$, resulting in 276 unique visibilities~\cite{thompson2017interferometry} per snapshot observation. The integration period is typically 1 second, and these visibilities are stored internally in registers. 

The correlator provides a Serial Peripheral Interface (SPI) interface~\cite{dhaker2018introduction}, and the registers are read by the control computer (see Section~\ref{sec:control-computer}). Various parameters of the telescope, such as integration time, are controlled by reading and writing the value of the registers in the correlator. 

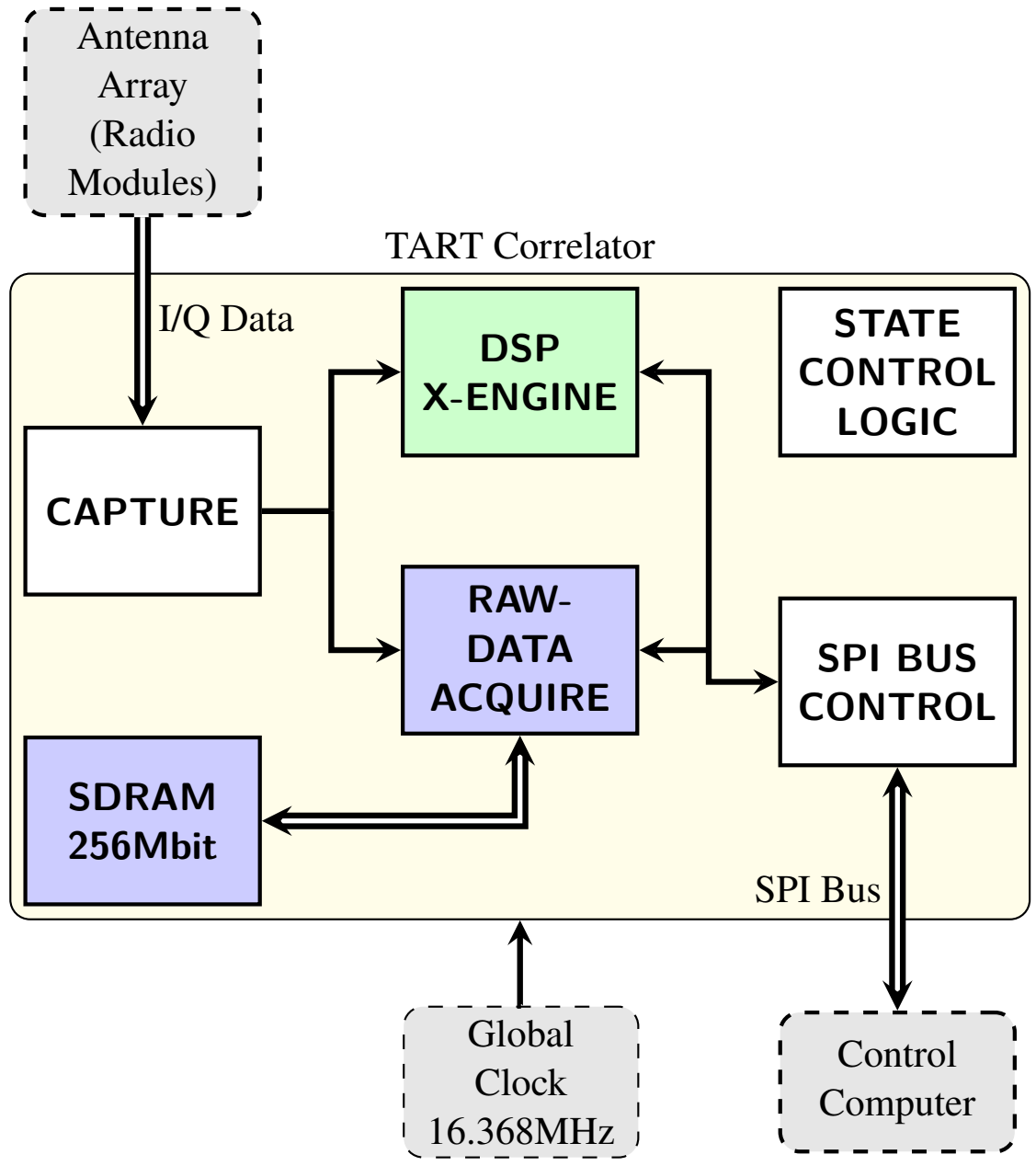
\begin{figure}
\begin{center}
\begin{tikzpicture}[
    node distance=1cm and 1cm, 
    auto,
    thick,
    box/.style={rectangle, draw, fill=white, text width=1.5cm, text centered, minimum height=1.2cm, font=\sffamily\bfseries},
    raw_mode/.style={rectangle, draw, fill=blue!20, text width=1.5cm, text centered, minimum height=1.2cm, font=\sffamily\bfseries},
    vis_mode/.style={rectangle, draw, fill=green!20, text width=1.5cm, text centered, minimum height=1.2cm, font=\sffamily\bfseries},
    wire/.style={-, >=stealth, very thick},
    arrow/.style={->, >=stealth, very thick},
    rarrow/.style={<-, >=stealth, very thick},
    bus/.style={->, >=stealth, very thick, double, double distance=1pt},
    spi_bus/.style={<->, >=stealth, very thick, double, double distance=1pt},
    biwire/.style={<->, >=stealth, very thick},
    external/.style={rectangle, draw, fill=gray!20, rounded corners, text width=1.5cm, text centered, minimum height=1cm, dashed},
    junction/.style={circle, fill=black, inner sep=1.5pt}
]

\node[box] (capture) {CAPTURE};
\node[external, above=of capture, yshift=0.5cm] (antennas) {Antenna Array (Radio Modules)};

\node[vis_mode, right=of capture, yshift=1cm] (dsp) {DSP X-ENGINE};
\node[raw_mode, right=of capture, yshift=-1cm] (acquire) {RAW-DATA ACQUIRE};

\node[box, right=of dsp] (control) {STATE CONTROL\\LOGIC};

\node[raw_mode, below=of capture] (sdram) {SDRAM\\256Mbit};
\node[box, below=of control] (spi_ctrl) {SPI BUS CONTROL};

\node[external, below=of spi_ctrl, yshift=-0.75cm] (host) {Control Computer};

\begin{pgfonlayer}{background}
    \node[fit=(capture)(dsp)(spi_ctrl)(acquire)(sdram)(control), draw, fill=yellow!10, rounded corners, label={above:TART Correlator}] (fpga) {};
\end{pgfonlayer}

\node[external, left=of host] (clock) {Global Clock 16.368MHz};


\draw[bus] (antennas) -- node[right] {I/Q Data} (capture);

\coordinate (cap_fork) at ($(capture.east)!0.5!(capture.east -| dsp.west)$);
\draw[wire] (capture) -- (cap_fork);
\draw[arrow] (cap_fork) |- (dsp);
\draw[arrow] (cap_fork) |- (acquire);

\coordinate (dsp_spi_junc) at ($(acquire.east)!0.5!(acquire.east -| spi_ctrl.west)$);

\draw[rarrow] (dsp.east) -| (dsp_spi_junc);
\draw[rarrow] (acquire.east) -| (dsp_spi_junc);
\draw[arrow] (dsp_spi_junc) |- (spi_ctrl);

\draw[spi_bus] (acquire.south) |- (sdram);

\draw[arrow] (clock.north) -| (fpga.south);
\draw[spi_bus] (spi_ctrl) -- node[left] {SPI Bus} (host);

\end{tikzpicture}
\caption{Block diagram of the TART-3 correlator. The CAPTURE module buffers the incoming data to avoid sampling close to global clock edges. The DSP X-ENGINE module performs real-time correlation and the RAW-DATA ACQUIRE and SDRAM modules manage the buffering and transfer of raw I/Q stored data. Blue-background modules are active only during raw-data mode, green-background modules are active only during visibility mode, and white-background modules are active in all modes. }
\label{fig:correlator_block_diagram}
\end{center}
\end{figure}

The correlator also contains a 256 MBit RAM buffer for storing raw I/Q samples from the radio telescope. This buffer can store two seconds of raw data from all antennas in the telescope. These data can be downloaded via the SPI interface; however, as the SPI interface speed is slower than the antenna data rate, raw-data download is not possible in real-time. The correlator code is written in the Verilog Hardware Description Language (HDL), and is available on github~\cite{Molteno_TART_Correlator_Code_2021}. Figure~\ref{fig:correlator_block_diagram} shows a block diagram of the TART-3 correlator.

\subsection{Control computer}
\label{sec:control-computer}

The TART telescope contains a control computer that is responsible for the configuration of and data collection from the radio telescope.  A web-based Application Programming Interface (API) allows control of telescope configuration and retrieving data -- either as visibilities, or raw receiver data. The control computer is a small single-board Raspberry-Pi 4 Model -B~\cite{raspberry_pi}.

\section{Software Architecture}
\label{sec:software}

The TART telescope software consists of several important but distinct categories. The first category is the {\em telescope API} that runs on the control computer. This software provides a means of sending and receiving TART data and modifying its operating parameters, and is described in Section~\ref{sec:telescope-api}. The second category of software is the user interface software for visualizing and controlling the TART. This user interface is described in Section~\ref{sec:telescope-user-interface} .

\subsection{The Telescope API}
\label{sec:telescope-api}
The telescope provides access to data and control over telescope operation using a RESTful Web-Service API~\cite{fielding2000architectural}. This API allows for the reading of data from anywhere using Hypertext Transfer Protocol (HTTP) requests. Although most API calls are available to the public, authentication is required for API endpoints that can change the configuration of the telescope.  The public facing components of the API allow TART data and images to be accessed freely from anywhere.

\subsection{Telescope Web Interface}
\label{sec:telescope-user-interface}

The TART telescope is typically operated via a browser through the telescope's web interface. Using this interface, the real-time live radio image can be viewed, videos recorded, cloud data retrieved, and raw and visibility data downloaded. 

\begin{figure*}
\includegraphics[width=\linewidth]{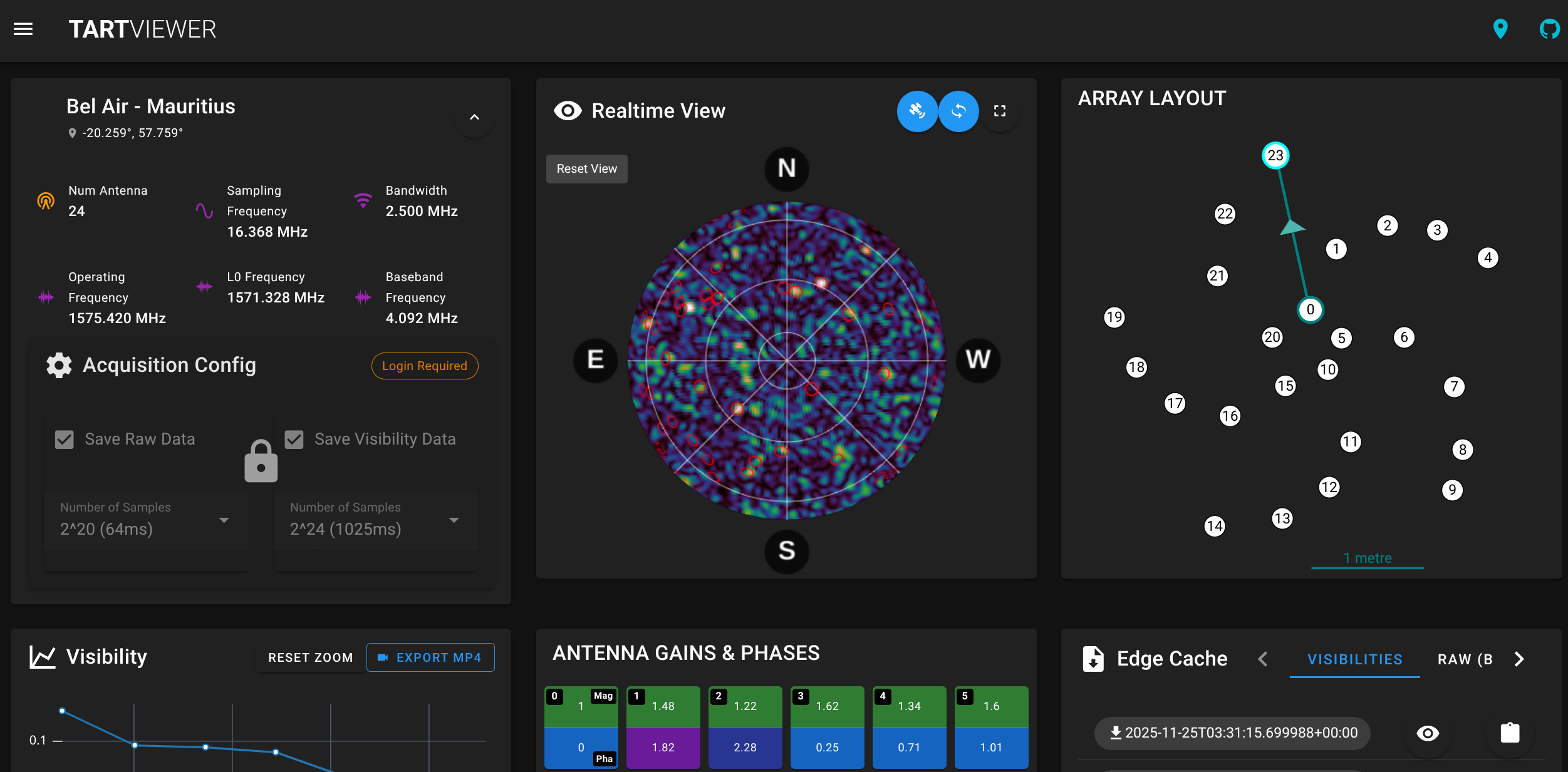}
\caption{The Web-based control interface for a TART telescope, which shows a live view of the radio sky above the TART location. The imaging is done in the user's browser. In this image the web interface is showing live images from the TART located at the Universit\'e des Mascareignes in Mauritius. }%
\label{fig:tart_web_app}
\end{figure*}

A web-based control interface that runs in a web browser and provides a convenient user interface to the RESTful API. This interface, shown in Figure~\ref{fig:tart_web_app}, provides real-time imaging, status information, and control over the telescope operating mode (allowing switching between `visibility' and `raw-data' mode). The interface allows monitoring of visibility phases and amplitudes, as well as calibration gains and phases, to provide condition monitoring of the telescope.

In addition to the web interface, the telescope RESTful API can be used with a wide variety of computer languages to automate data analysis. Figure~\ref{fig:tart_web_app} shows the telescope web interface connected to the TART located at the Universit\'e des Mascareignes in Mauritius.

\subsection{TART tools}

The TART software includes tools (tart2ms~\cite{Molteno_TART2MS_tool_to_2025}) to convert downloaded TART data to standard measurement set (MS) format~\cite{kemball2000measurementset}. Once in MS form, data can be processed using standard radio astronomical frameworks such as CASA~\cite{casa2022casa} or Stimela~\cite{SMIRNOV2025100959}, and imaged with any standard aperture-synthesis imaging package such as WSClean~\cite{offringa2014wsclean}.

\subsection{Calibration Software}

The GNSS satellites that are visible with the TART, have very precisely known positions and signal structure. These are ideal sources for calibrating the gains and phases of the antenna array. The TART source catalogue provides satellite positions and is available using a RESTful API. This catalogue server is part of the TART telescope software~\cite{Molteno_Catalogue_2025}. 

There are two calibration approaches, the first is automatic continuous calibration performed regularly on a compute cluster, and the second is performed as part of an imaging pipeline. Both calibration processes make use of data downloaded from the TART, and known source positions to estimate the calibration gains and phases. 

The continuous calibration process solves for calibration gains and phases from each antenna. The core calibration algorithm minimizes an objective function across a 47-dimensional parameter space—consisting of a global coordinate rotation angle alongside real and imaginary complex gain components for 23 antennas relative to a reference element. The process generates a synthesized sky image from measured baseline visibilities via inverse Fast Fourier Transforms (IFFT) and calculates a cost score based on the image's signal-to-noise ratio combined with the image intensity at known satellite positions. To navigate the non-linear objective function and avoid getting trapped in local minima, the system utilizes a Basin-Hopping stochastic optimization algorithm, which pairs Monte random parameter perturbations with Quasi-Newton steepest-descent local minimizations~\cite{wales1997global}. After an initial full global search locates the optimal baseline parameters (requiring roughly 250,000 function evaluations), the system transitions to computationally lightweight steepest-descent updates during normal operation to continuously track temperature drifts and phase variations.  This calibration procedure is described in more detail in~\cite{continuous_TART_cal}.

\subsection{Imaging Pipelines}

Although imaging can be performed through the web interface (see Section~\ref{sec:telescope-user-interface}), TART data can be converted to measurement sets and processed using traditional radio astronomy imaging and calibration pipelines. Learning imaging on TART using conventional pipelines has the advantage of teaching a framework for working with TART data that can be transferred to other radio telescope data. 

The TART software includes specialized wide-field imaging algorithms that image in spherical coordinates on the half-sphere (see, for example,~\cite{Molteno_Discrete_Sky_Operator_2023}) that can also be used. These TART tools are available as cabs for the Stimela2~\cite{SMIRNOV2025100959} radio astronomy pipeline framework, as the  tart\_cargo~\cite{Molteno_TART_CARGO_Stimela_Cabs_2025} python package. 
Section~\ref{sec:tart_imaging_results} shows a stimela-based TART imaging pipeline, and the corresponding images of the radio sky above Ghana.

\section{Results: TART imaging}
\label{sec:tart_imaging_results}

\begin{figure}
     \centering
     \begin{subfigure}[t]{0.3\textheight}
         \centering
         \includegraphics[width=\textwidth]{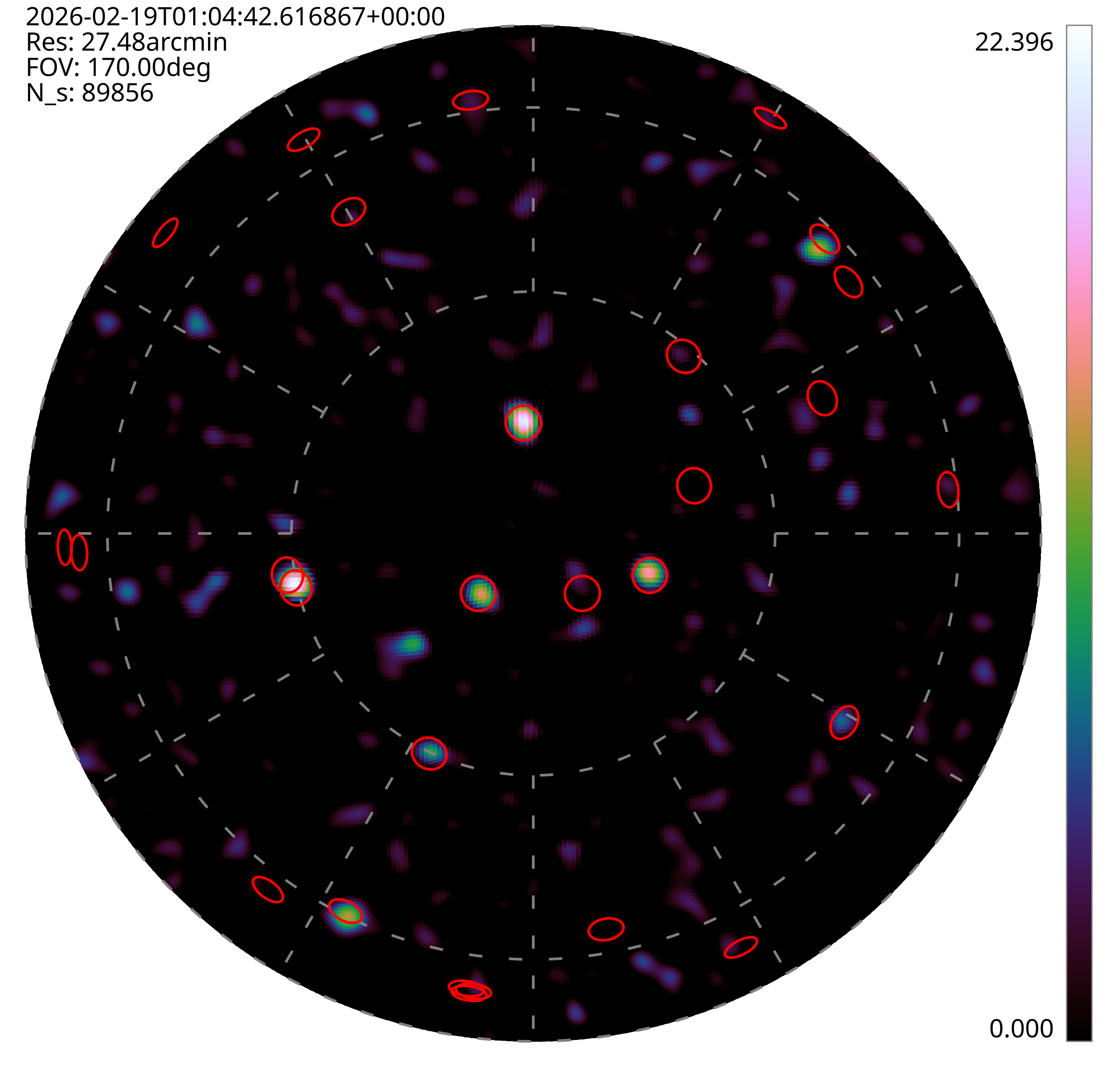}
         \caption{Regularized least squares (DiSkO) image. }
         \label{fig:tart_disko}
     \end{subfigure}
     \hfill
     \begin{subfigure}[t]{0.3\textheight}
         \centering
         \includegraphics[width=\textwidth]{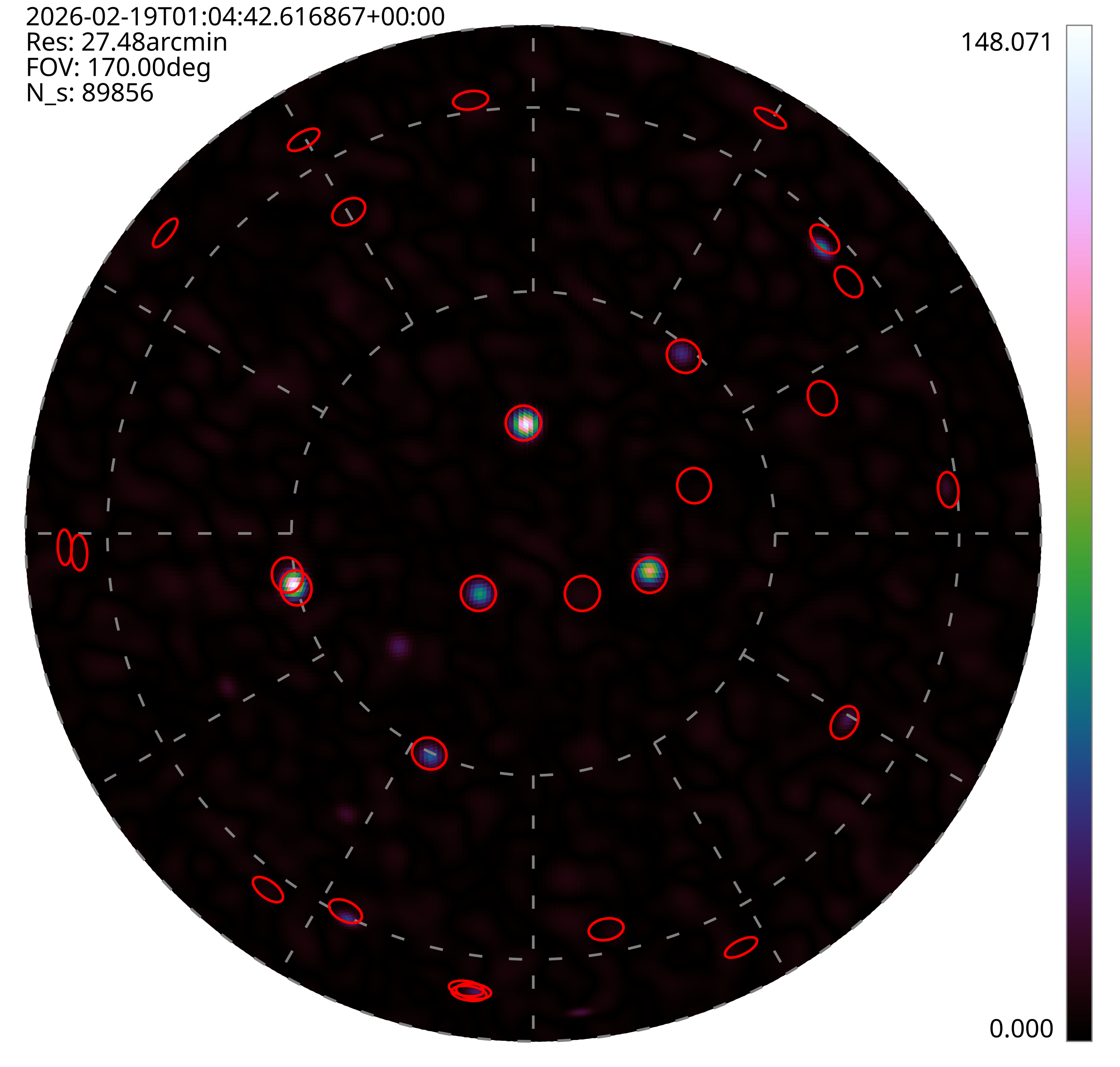}
         \caption{Model-based visibility-space deconvolution (Spotless) image.}
         \label{fig:tart_spotless}
     \end{subfigure}
     \hfill
     \begin{subfigure}[t]{0.3\textheight}
         \centering
         \includegraphics[width=\textwidth]{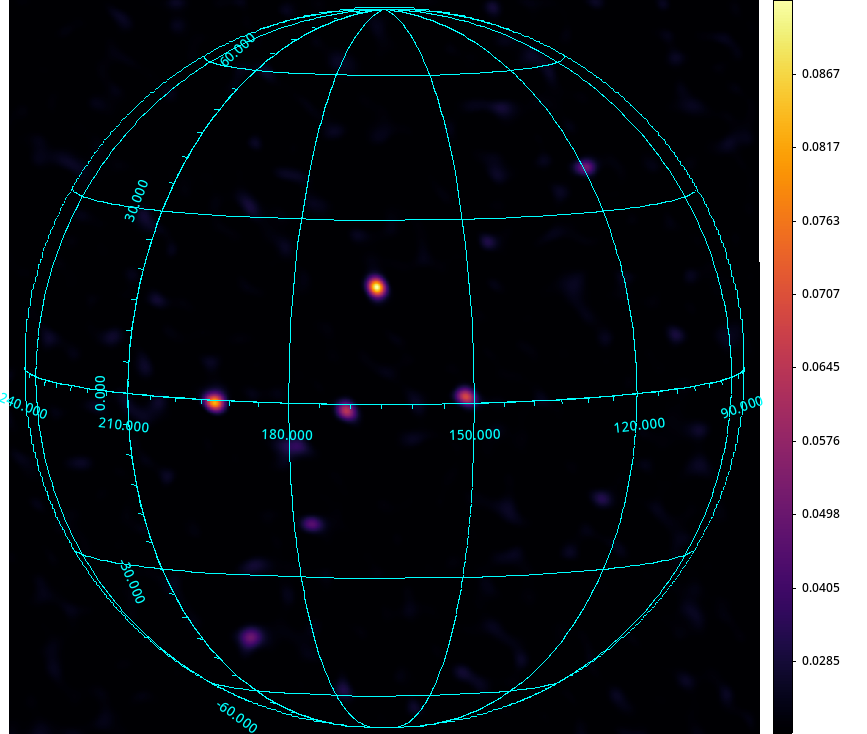}
         \caption{Restored CLEAN image from the same TART measurement set.}
         \label{fig:tart_wsclean}
     \end{subfigure}
    \caption{Images generated from a single snapshot of visibilities from the TART located in Ghana. Known sources are shown as red circles. The top image is generated using DiSkO, a regularized least-squares imager. The bottom image is generated using Spotless, a model-based imager performed in visibility space. In \ref{fig:tart_disko} and~\ref{fig:tart_spotless}, the image is normalized to the RMS noise level.}
    \label{fig:tart_images}
\end{figure}

The Stimela imaging pipeline shown below automatically downloads data from the TART telescope, converts it to a measurement set, performs standard CLEAN imaging, and then generates two TART specific algorithms on the same data. This produces the images shown in Figure~\ref{fig:tart_images}.

This, and other stimela recipes for TART imaging and calibration, including the raw data used, are available online~\cite{Molteno_TART_radio_imaging_2026}.

\begin{lstlisting}[language=stimela]
# Basic TART Stimela 2.0 processing pipeline

_include:
    - (tartcargo):
        - tart2ms.yml
        - tart-download-data.yml
        - disko.yml
        - spotless.yml
        - disko-draw.yml
    - (cultcargo):
        - wsclean.yml

tart-image:
    info: TART imaging using stimela

    inputs:
        tart:
            dtype: str
            required: true
            info: "TART name (e.g. mu-udm)"

    outputs:
        ms:
            dtype: MS
            default:  '{current.tart}.ms'
        hdf:
            dtype: File
            default: '{current.tart}-vis.hdf'
        svg:
            dtype: File
            default: '{current.tart}.svg'
        imgdir:
          dtype: Directory
          mkdir: true
          default: img

    assign:
        api: 'https://api.elec.ac.nz/tart/{recipe.tart}'

    steps:
        download-hdf:
            cab: tart-download-data
            skip: =EXISTS(recipe.hdf)
            params:
                api: =recipe.api
                vis: true
                file: =recipe.hdf

        create-ms:
            cab: tart2ms
            skip: =EXISTS(recipe.ms)
            params:
                hdf: =steps.download-hdf.file
                ms: =recipe.ms
                clobber: true
                single-field: true
                rephase: obs-midpoint

        wsclean-image:
            cab: wsclean
            params:
                ms: =recipe.ms
                prefix: '{recipe.imgdir}/{recipe.tart}'
                temp-dir: '.'
                size: 1024
                scale: 7.0amin
                column: DATA
                nchan: 1
                pol: RR
                weight: briggs 0.0
                intervals-out: 1
                gain: 0.1
                mgain: 0.85
                niter: 5000
                auto-mask: 1.5
                auto-threshold: 1
                no-update-model-required: true
                no-fast-subminor: true
                no-min-grid-resolution: true


        spotless-image:
            cab: spotless
            params:
                ms: =recipe.ms
                HDF: ='{recipe.tart}-spotless.sphere'
                nvis: 10000
                multimodel: true
                healpix: true
                fov: 170deg
                res: 0.5deg

        spotless-draw:
            cab: disko-draw
            params:
                hdf: ='{recipe.tart}-spotless.sphere'
                show-sources: true
                scale-mad: true
                SVG: ='spotless-{recipe.svg}'

        disko-image:
            cab: disko
            params:
                ms: =recipe.ms
                HDF: ='{recipe.tart}-disko.sphere'
                nvis: 10000
                healpix: true
                fov: 170deg
                res: 0.5deg
                lasso: true
                alpha: 0.006
                l1-ratio: 0.02

        disko-draw:
            cab: disko-draw
            params:
                hdf: ='{recipe.tart}-disko.sphere'
                show-sources: true
                scale-mad: true
                SVG: ='disko-{recipe.svg}'
\end{lstlisting}

In a TART with all antennas functioning and with typical environmental conditions, the noise level in the images is typically $\sim$100 times lower than the peak intensity of the point-source from a GNSS satellite. Figure~\ref{fig:tart_images} shows an example where the S/N is 148).

As the Spectral flux density of GNSS source signals has been well characterized, we can use a conservative estimate of $\sim 100$ for the signal-to-noise level, and characterize the telescope sensitivity. This is described in the next section.

\subsection{Sensitivity Estimates}

Sensitivity estimates for the TART telescope are complicated by the automatic gain control (AGC) circuitry built into the integrated GNSS radio receiver~\cite{max2769b:datasheet}. The AGC circuit dynamically adjusts the receiver gain to maintain a constant amplitude of the received signal. For wide-beam antennas such as those in the TART, the received signal is typically dominated by thermal noise due to the presence of objects at ambient temperature within the antenna beam~\cite{borre2007software}. Fortunately, this thermal noise is not correlated between receivers, and so does not limit imaging from visibilities.  The AGC gain is not known a-priori, and can make flux calibration difficult. However, for the TART operating in GNSS bands, the sensitivity of the receiver can be estimated after imaging by comparing the measured flux of sources with known flux with the imaging noise level.

Using published plots of Spectral Flux Densities (SFD) from ground station~\cite{steigenberger2018flex, thoelert2019signal}, measurements it is possible to fit a modeled spectrum and estimate the integrated spectral flux density over the band-pass response of the TART telescopes (see Figure~\ref{fig:rf_bandpass_response}). 

\begin{figure}
\centering
    \begin{subfigure}[t]{0.49\linewidth}
        \includegraphics[width=\linewidth]{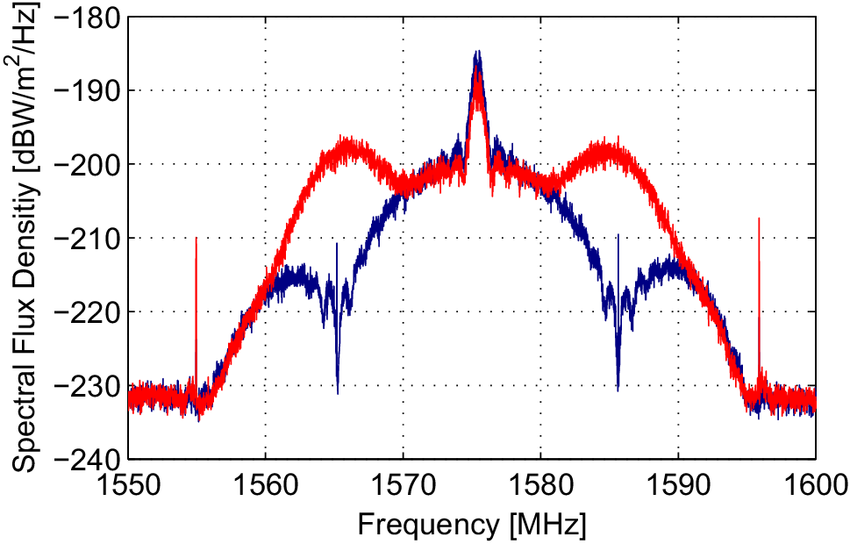}
        \caption{Spectral flux density of the GPS Block II satellite measured by the 30m telescope at DLR~\cite{steigenberger2018flex}.}
        \label{fig:measured_flux_density}
    \end{subfigure} \hfill
    \begin{subfigure}[t]{0.49\linewidth}
        \includegraphics[width=\linewidth]{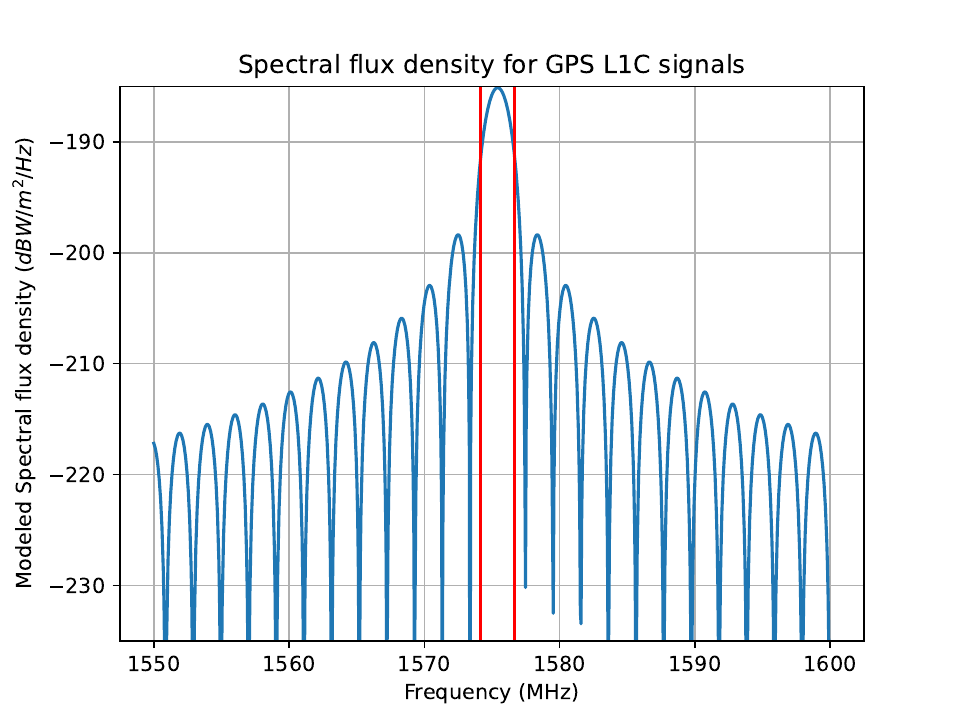}
        \caption{Modeled BPSK spectrum for the L1/C GPS signal shown in (a). The central peak lies within the bandwidth of the TART receiver (red lines)}
        \label{fig:modeled_flux_density}
    \end{subfigure}
    \caption{Measured and modeled spectral flux density. Fitting the BPSK model to the measured data allows us to integrate over the TART bandwidth to estimate the average source power in the receiver band as $\sim$60,000 Jy.}
    \label{fig:spectral_flux_density}
\end{figure}

The SFD $S(\nu)$ of GPS L1C signals as a function of frequency $\nu$, is well modeled as that of a Binary Phase Shift Key (BPSK) signal with a random sequence of binary values~\cite{borre2007software}. The functional form of the SFD is~\cite{lathi1998modern},
\begin{equation} S_{BPSK}(\nu) = \frac{A^2 T_b}{4} \left\{ \text{sinc}^2\left[ T_b(\nu - \nu_c) \right] + \text{sinc}^2\left[ T_b(\nu + \nu_c) \right] \right\},
\end{equation}
where $\nu$ is the frequency, $\nu_c$ is the center-frequency of the GPS L1 band, and $T_b$ is the time per bit of the BPSK modulation ($\sim 977 \unit{\micro\second}$) and $A$ the carrier amplitude. The maximum value of the SFD, $S_{max}$ is $\frac{A^2 T_b}{4}$. Figure~\ref{fig:modeled_flux_density} shows this modeled SFD normalized to $S_{max} = 1$.

\begin{figure}
    \includegraphics[width=\linewidth]{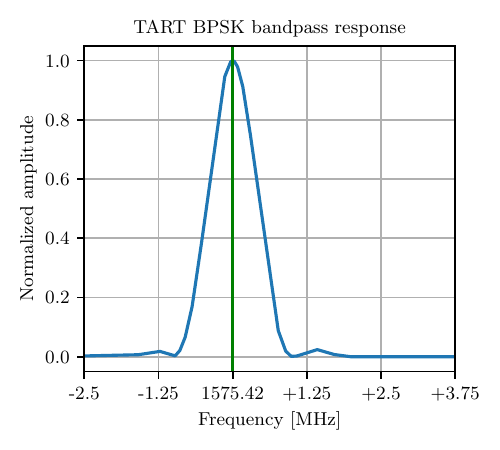}
    \caption{Bandpass-filtered BPSK spectrum for the TART receiver. We can estimate the band-filling factor of a GNSS signal from the mean value of this function over the TART band ($\pm 1.25 \unit{\mega\hertz}$) either side of the center frequency.}
    \label{fig:bpsk_bandpass_response}
\end{figure}

Ground based SFD measurements(see Figure~\ref{fig:measured_flux_density}) show that the maximum GPS SFD measured is -188 \unit{\decibel\watt\per\square\meter\per\hertz}, which is equivalent to $15.8 \times 10^{-19}$ \unit{\watt\per\square\meter\per\hertz} or 15.8 mega-Jansky. From these measurements, we can estimate the integrated flux over the TART band by multiplying the bandpass response by the BPSK spectral flux density. This is shown in Figure~\ref{fig:bpsk_bandpass_response}, which shows that, across the TART band ($1575.42 \pm 1.25 \unit{\mega\hertz}$), the average SFD is approximately 33\% of the maximum value, or 5.2 MJy.

Using an estimate of the imaging signal-to-noise ratio of approximately 100 (from Figure~\ref{fig:tart_images}), the sensitivity of the TART telescope is estimated to be $\sim 52$ \unit{\kilo\Jansky}.

\section{Discussion \& Conclusions}
\label{sec:discussion}

We have described the architecture and operation of the TART radio telescope, and estimated it's sensitivity to be approximately 50\unit{\kilo\Jansky}. To date, TART has been installed in New Zealand, South Africa, Mauritius, Kenya, Botswana, Ghana~\cite{proven2025training}, Namibia and Bangladesh as a platform for the development of algorithms, as well as research, teaching, and outreach. The majority of these installations are connected to a cloud based network that provides public access to all telescope data (see Figure~\ref{fig:tart_web_map}).

Several improvements are planned. The software, both for imaging and for operation, is undergoing continuous development, and several hardware improvements are underway to improve sensitivity. These include new antenna designs with more uniform beam patterns -- existing electrically-small antennas have significant variability in their beams, causing reduced imaging sensitivity. A new correlator is under development that will allow the TART to use all radio slots on the motherboard, increasing the number of antennas to 32 and significantly improving the number of baselines and sensitivity. The third hardware improvement is an improved radio design with bandwidth increasing from 2.5 MHz to approximately 26 MHz, once again increasing sensitivity (although increasing correlator complexity as the data will require channelization, tentatively into 8 channels).

In addition to hardware improvements, a project is underway to combine signals from multiple TART telescopes using Very Long Baseline Interferometry (VLBI) techniques with clock recovery from the GNSS time signals embedded in the GNSS signals. Early results are promising with fringes observed between bright GNSS satellite targets between Namibia and South Africa over a baseline of ~2000 km. These fringes can be observed between single antenna observations on each of the TARTs. Another planned modification is to implement a large ~(10 minute) buffer of raw data on each telescope. VLBI would then be used to refine the positions of bright transient sources after local transient event detection using this buffer of raw data.

Finally, we plan to use parallax between close ($~\sim 10$km) separation pairs of TART telescopes to try and estimate the distance to transient events that occur in the upper atmosphere and in low-medium earth orbit. Pairs of TART also enable positive discrimination between transient events and local radio frequency interference.

\section*{Acknowledgments}

The authors acknowledge the work of the many contributors to the TART project. These include Mike Fields (HamsterNZ) who helped with memory interfaces for the early versions of the telescope correlator, Rikus Human who helped with antenna array design, and Charles Shaw who worked on an early prototype. We also acknowledge the many people who have helped with TART workshops, these include Prof Oleg Smirnov (Rhodes University, SARAO), Prof Melvin Hoare (University of Leeds, DARA), Dr Landman Bester (SARAO), Dr Nadeem Oozeer (SARAO), Rikus Human (Alphawave), Dr Khan Asad (IUB) and Shaoib Mirza (IUB). In addition the TART team would like to thank the South African Radio Astronomy Observatory (SARAO) and the Development in Africa with Radio Astronomy (DARA)  project for their invaluable support for the deployment of TART telescopes in Africa.

\section*{Data Availability}

All data used in this paper were downloaded from the public TART telescope at \url{https://tart.elec.ac.nz/viewer/ghana}. A repository of the data used in this report is also available~\cite{Molteno_TART_radio_imaging_2026}.




\printbibliography




\bsp	
\label{lastpage}
\end{document}